\documentclass[twocolumn,superscriptaddress,PRB]{revtex4}

\usepackage{amsmath}
\usepackage{amssymb}
\usepackage{graphicx}
\usepackage{color}
\usepackage{bm}
\usepackage{enumerate}

\begin{document}

\title{Evidence for time-reversal symmetry breaking of the superconducting state near twin-boundary interfaces in FeSe}

\author{T.~Watashige}
\affiliation{Department of Physics, Kyoto University, Kyoto 606-8502, Japan}
\affiliation{RIKEN Center for Emergent Matter Science, Wako, Saitama 351-0198, Japan}

\author{Y.~Tsutsumi}
\affiliation{Condensed Matter Theory Laboratory, RIKEN, Wako, Saitama 351-0198, Japan}

\author{T.~Hanaguri}
\affiliation{RIKEN Center for Emergent Matter Science, Wako, Saitama 351-0198, Japan}

\author{Y.~Kohsaka}
\affiliation{RIKEN Center for Emergent Matter Science, Wako, Saitama 351-0198, Japan}

\author{S.~Kasahara}
\affiliation{Department of Physics, Kyoto University, Kyoto 606-8502, Japan}

\author{A.~Furusaki}
\affiliation{RIKEN Center for Emergent Matter Science, Wako, Saitama 351-0198, Japan}
\affiliation{Condensed Matter Theory Laboratory, RIKEN, Wako, Saitama 351-0198, Japan}

\author{M.~Sigrist}
\affiliation{Theoretische Physik, ETH Z\"urich, CH-8093 Z\"urich, Switzerland}

\author{C.~Meingast}
\affiliation{Institute of Solid State Physics (IFP), Karlsruhe Institute of Technology, D-76021 Karlsruhe, Germany}

\author{T.~Wolf}
\affiliation{Institute of Solid State Physics (IFP), Karlsruhe Institute of Technology, D-76021 Karlsruhe, Germany}

\author{H.~v. L\"ohneysen}
\affiliation{Institute of Solid State Physics (IFP), Karlsruhe Institute of Technology, D-76021 Karlsruhe, Germany}

\author{T.~Shibauchi}
\affiliation{Department of Advanced Materials Science, University of Tokyo, Chiba 277-8561, Japan}
\affiliation{Department of Physics, Kyoto University, Kyoto 606-8502, Japan}

\author{Y.~Matsuda}
\affiliation{Department of Physics, Kyoto University, Kyoto 606-8502, Japan}

\date{\today}

\begin{abstract}
{
Junctions and interfaces consisting of unconventional superconductors provide an excellent experimental playground to study exotic phenomena related to the phase of the order parameter.
Not only the complex structure of unconventional order parameters have an impact on the Josephson effects, but also may profoundly alter the quasi-particle excitation spectrum near a junction.
Here, by using spectroscopic-imaging scanning tunneling microscopy, we visualize the spatial evolution of the local density of states (LDOS) near twin boundaries (TBs) of the nodal superconductor FeSe. 
The $\pi/2$ rotation of the crystallographic orientation across the TB twists the structure of the unconventional order parameter, which may, in principle, bring about a zero-energy LDOS peak at the TB.
The LDOS at the TB observed in our study, in contrast, does not exhibit any signature of a zero-energy peak and an apparent gap amplitude remains finite all the way across the TB.
The low-energy quasiparticle excitations associated with the gap nodes are affected by the TB over a distance more than an order of magnitude larger than the coherence length $\xi_{ab}$.
The modification of the low-energy states is even more prominent in the region between two neighboring TBs separated by a distance $\approx7\xi_{ab}$.
In this region the spectral weight near the Fermi level ($\approx\pm$0.2~meV) due to the nodal quasiparticle spectrum is almost completely removed.
These behaviors suggest that the TB induces a fully-gapped state, invoking a possible twist of the order parameter structure which breaks time-reversal symmetry.
}
\end{abstract}

\maketitle

\section{Introduction}

When two superconductors are in close proximity, they are influenced by each other via the tunneling of Cooper pairs.
The Cooper-pair tunneling results in the flow of a superconducting Josephson current, which has been studied for decades and is used in various superconducting quantum devices~\cite{Duzer_book}.
The Josephson current is governed by the phase difference of the order parameters of the two superconductors.
Therefore, Josephson junctions consisting of unconventional superconductors, where the superconducting order parameter changes its sign depending on the momentum direction, serve as a unique platform where novel phase-related phenomena, e.g., spontaneous formation of half flux quanta in a tri-junction of cuprate superconductors~\cite{Tsui2000RMP}, take place.
Compared with the well-investigated Josephson currents, the spatial and energy dependence of the superconducting order parameter and quasiparticle states around these junctions remain to be understood.

Recent progress in scanning tunneling microscopy (STM) and spectroscopy (STS) technologies opens up a way to directly visualize the spatial variation of the electronic states in superconducting hetero-structures~\cite{Kim2012NatPhys,Serrier-Garcia2013PRL,Cherkez2014PRX}.
However, STM/STS studies on superconducting junctions made of unconventional superconductors are still demanding.
There are two reasons which make it difficult to study unconventional junctions.
First, it is often challenging to artificially fabricate well-defined junctions of unconventional superconductors.
Second, in most of unconventional superconductors, surfaces are not electronically neutral; the resultant charge accumulation at the surfaces prevents STM/STS from accessing bulk superconducting properties.
In this study, we solve these problems by inspecting the twin boundaries (TBs) in the nodal iron-based superconductor FeSe~\cite{Song2011Science,Kasahara2014PNAS}.

The TB is a crystallographic plane in a crystal shared by two neighboring domains with one being the mirror image of the other.
The TBs are often formed by a tetragonal-to-orthorhombic structural phase transition, which reduces the four-fold ($C_4$) symmetry at high temperature to two-fold ($C_2$) symmetry at low temperature.
In such a case, the orthorhombic crystal may contain the TBs parallel to the (110) plane, which act as an atomically well-defined junction.
Some unconventional-superconductor-related materials, such as YBa$_2$Cu$_3$O$_{7-\delta}$, AE(Fe$_{1-x}$Co$_x$)$_2$As$_2$ (AE: alkali-earth element) and NaFeAs, do form TBs upon the tetragonal-to-orthorhombic transition which were identified by STM/STS measurements~\cite{Derro2000PhysicaC,Chuang2010Science,Rosenthal2014NatPhys}.
However, unavoidable surface state formation and/or insufficient amount of chemical doping prevent the STM/STS measurements to access superconductivity near TBs in these materials.

FeSe (superconducting transition temperature $T_c\approx9$~K~\cite{Hsu2008PNAS}) is a promising candidate for studying the effects of TBs on unconventional superconductivity by STM/STS.
Among various iron-based superconductors, FeSe has the simplest crystal structure [Fig.~1(a)] in which electronically neutral two-dimensional FeSe layers are stacked along the $c$ axis~\cite{Hsu2008PNAS}.
This guarantees the perfect cleaved surface which is electronically neutral.
The tetragonal-to-orthorhombic structural phase transition, which is likely caused by the orbital ordering~\cite{Nakayama2014,Shimojima2014PRB,Watson2015condmat_a,Zhang2015condmat,Boehmer2015PRL,Baek2015NatMat}, occurs at $T_s\approx90$~K and the TBs are spontaneously formed in the orthorhombic phase as illustrated in Fig.~1(b).

Band-structure calculations show that the Fermi surface of FeSe consists of hole cylinders around the zone center and compensating electron cylinders around the zone corner~\cite{Subedi2008PRB,Aichhorn2010PRB}.
Several measurements, including penetration depth, quasiparticle interference, thermoelectric response~\cite{Kasahara2014PNAS}, quantum oscillations~\cite{Terashima2014PRB,Watson2015condmat_a,Watson2015condmat_b}, and angle-resolved photoemission spectroscopy (ARPES)~\cite{Nakayama2014,Shimojima2014PRB,Watson2015condmat_a,Zhang2015condmat,Maletz2014PRB} reveal that the Fermi surface in the orthorhombic phase consists of one hole and one (or two \cite{Watson2015condmat_a,Watson2015condmat_b}) electron bands, both of which have very low carrier densities.
The tunneling spectrum~\cite{Song2011Science}, temperature dependence of the penetration depth down to 80~mK and the residual thermal conductivity at $T\rightarrow 0$~\cite{Kasahara2014PNAS}, all provide strong evidence that FeSe is an unconventional superconductor with line nodes in the superconducting gap.

The TBs in FeSe have been studied by low-temperature (4.2~K) STM/STS in the films grown by molecular beam epitaxy and the suppression of superconductivity by the TBs has been reported~\cite{Song2012PRL}.
We performed STM/STS measurements at much lower temperature ($\approx0.4$~K) in vapor-grown single crystals to examine the details of the superconducting gap and quasiparticle excitations near the TBs.

\section{Experimental method}

STM/STS experiments were conducted in a constant-current mode with a commercial ultra-high vacuum very-low temperature STM (UNISOKU, USM-1300) modified by ourselves~\cite{Hanaguri2006JPhys}.
The samples used in this study were high-quality bulk single crystals grown using the vapor transport method~\cite{Boehmer2013PRB}.
Superconducting transition temperature defined at zero resistance is about 9~K.
These crystals are undoped and stoichiometric, enabling us to investigate uniform and clean TBs.
Samples were cleaved {\it in-situ} at liquid N$_2$ temperature to prepare clean and flat (001) surfaces.
Immediately after cleaving, the samples were transferred to the STM unit kept below 10~K.
We used electrochemically-etched polycrystalline tungsten wires for the scanning tips which were cleaned and sharpened {\it in-situ} by field evaporation using field-ion microscopy.
The tunneling conductance $g(\bm{r}, E) \equiv dI_t/dV_s(\bm{r}, E)$ reflecting the local density of states (LDOS) at a position $\bm{r}$ and energy $E$, was acquired by standard lock-in technique.
Here, $I_t$ and $V_s$ denote the tunneling current and the sample-bias voltage, respectively.

\section{Results and discussion}

\subsection{Imaging the twin boundary in FeSe}

\begin{figure*}
\includegraphics[width=130mm]{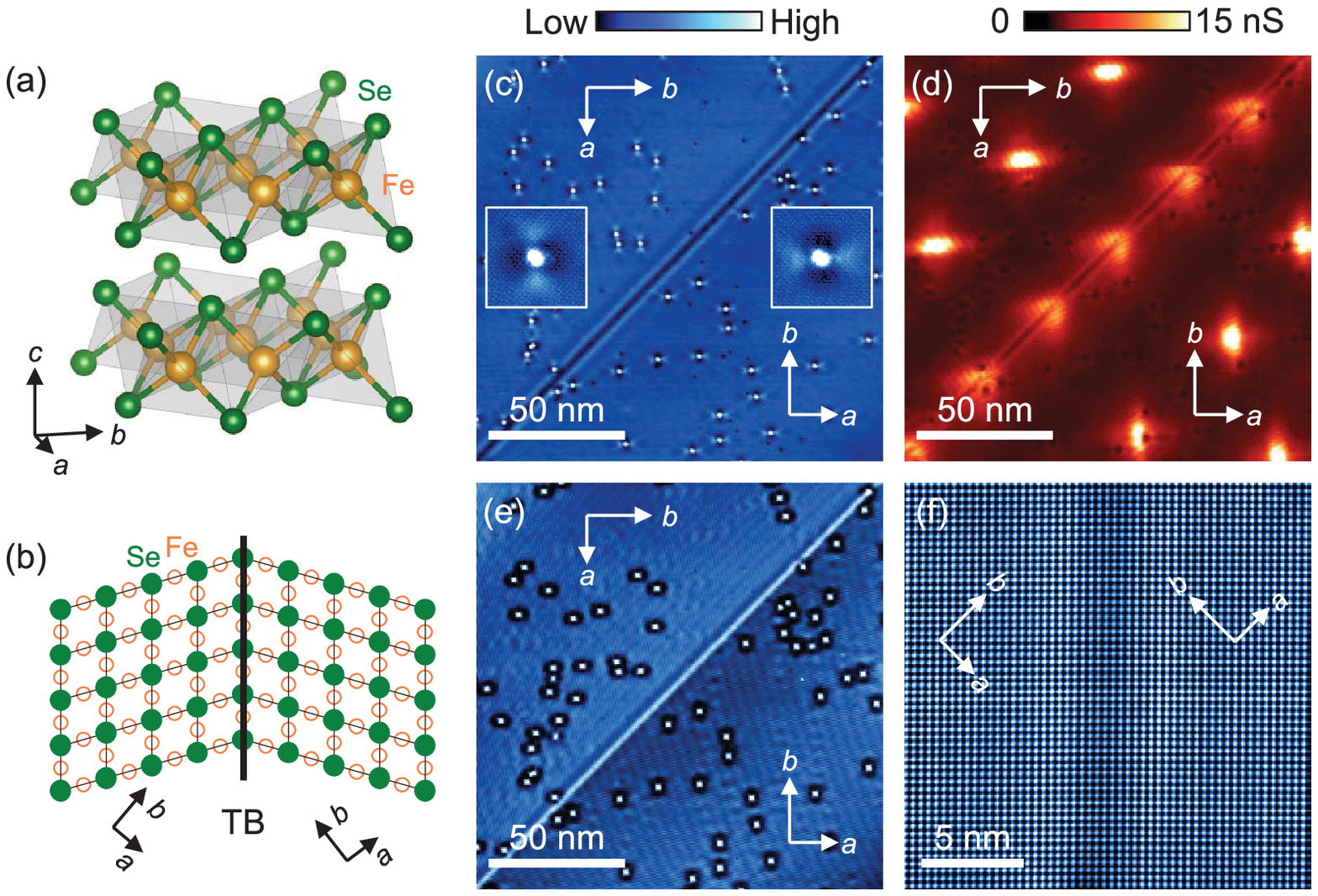}
\caption{
(a)
Crystal structure of FeSe visualized using the VESTA program~\cite{Monma2011JAC}.
(b)
Schematic top view of the atomic arrangement near the TB of FeSe (not in scale).
Green filled circle and orange open circle denote top-most Se and Fe atoms, respectively.
Se atoms beneath the Fe layer are not shown.
(c)
A constant-current STM image of the cleaved (001) surface of FeSe at 1.5~K showing the TB running from bottom left to top right.
Crystallographic axes parallel to the Fe-Fe direction are shown by white arrows ($b>a$).
The two insets show a magnified image of the defect (8.8~nm $\times$ 8.8~nm) in the upper-left or lower-right domain.
Note that the pattern is rotated by $\pi/2$ between the two domains.
The set-up conditions for imaging were $V_s=+95$~mV and $I_t=10$~pA.
(d)
Zero-bias conductance image $g(\bm{r}, E=0)$ at 1.5~K showing vortices.
A magnetic field of 1~T was applied along the $c$ axis.
The tip was stabilized at $V_s=+10$~mV and $I_t=100$~pA.
A bias modulation amplitude $V_{\rm mod}=0.21$~mV$_{\rm rms}$ was used for spectroscopy.
(e)
A low-bias STM image at 1.5~K taken with $V_s=+20$~mV and $I_t=10$~pA.
The field of view for (c)-(e) is the same.
(f)
An atomic-resolution STM image near the TB which is running vertically in the center of the field of view.
$V_s=+95$~mV and $I_t=100$~pA.
}
\end{figure*}

Figure~1(c) depicts an STM image of the cleaved surface of an FeSe single crystal at temperature $T=1.5$~K.
The image demonstrates the extremely small concentration of defects, i.e. about one defect per 5000 Fe atoms in the (001) plane.
There is a shallow ``groove'' running along the [110] direction of the Fe lattice across which the unidirectional feature around the point defect is rotated by $\pi/2$, indicating that the ``groove'' represents the TB.
We also observed that the elongated vortex cores~\cite{Song2011Science}, which were imaged by mapping $g(\bm{r}, E=0)$ in a magnetic field, are rotated by $\pi/2$ across the TB [Fig.~1(d)].
What is intriguing is that the vortices trapped at the TB are not elongated along the TB, demonstrating that the critical current density across the TB is comparable to that of the bulk.
The STM image of the TB at a lower bias voltage is not a ``groove'' but a ``ridge'' [Figs.~1(e)].
This suggests that the apparent corrugations near the TB are primarily associated with the electronic-state variations; the actual surface topography near the TB may be essentially flat.
A magnified STM image near the TB is shown in Fig.~1(f).
A regular square lattice of the top-most Se atoms is well maintained even in the close vicinity of the TB.
These observations indicate that the TB in FeSe is an atomically sharp superconducting junction with minimal strain to the lattice.
(A detailed argument regarding the absence of strain is given in Appendix~A.)

\subsection{Local density of states across the twin boundaries}

\begin{figure*}
\includegraphics[width=120mm]{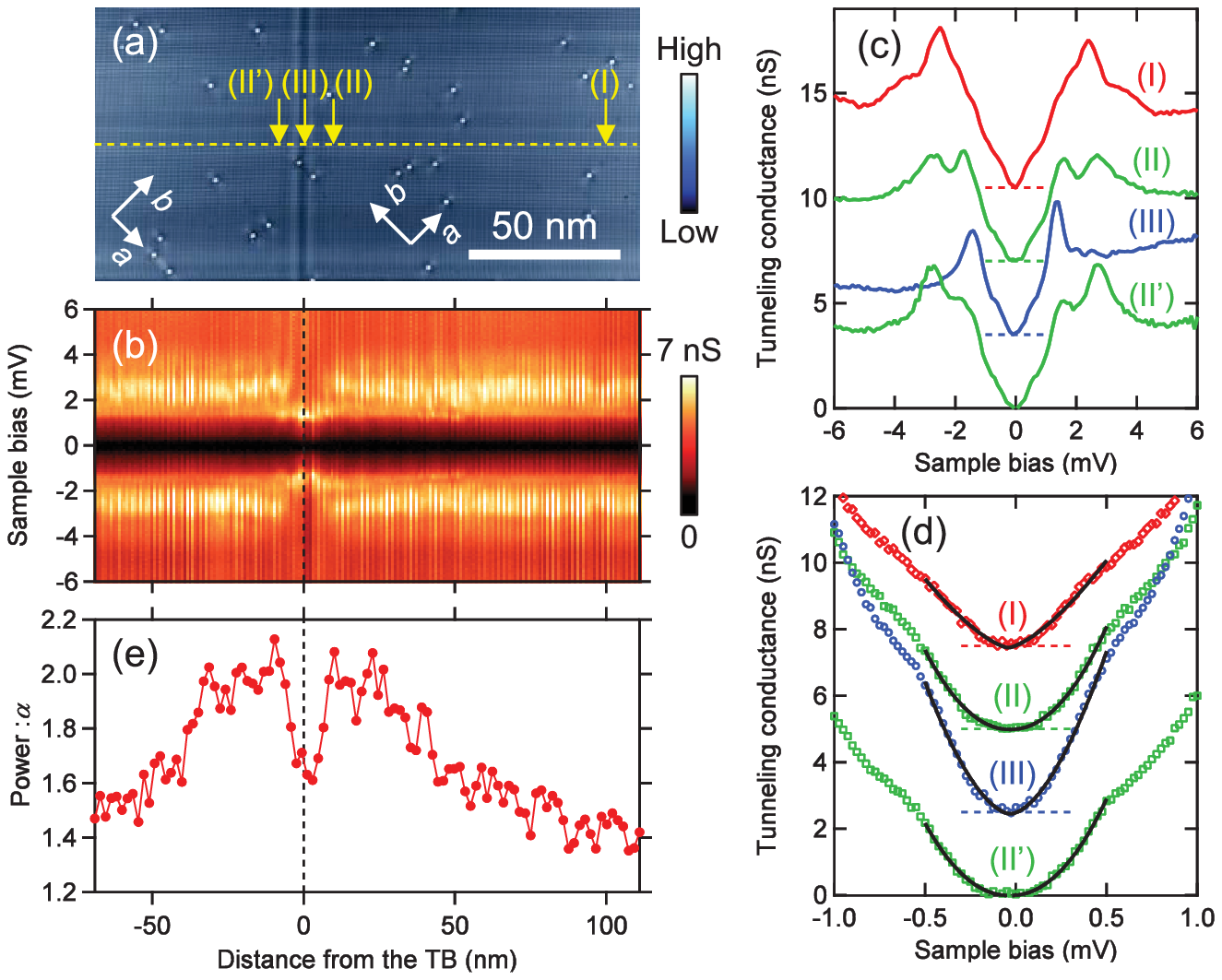}
\caption{
(a)
A constant-current STM image near a TB taken at $V_s=+95$~mV and $I_t=100$~pA.
(b)
Intensity plot of $g(\bm{r}, E)$ along the yellow broken line in (a).
$V_s=+20$~mV, $I_t=100$~pA and $V_{\rm mod}=0.05$~mV$_{\rm rms}$.
(c)
Tunneling spectra at the representative points indicated in (a).
Positions (II) and (II') are symmetric about the TB.
$V_s=+20$~mV, $I_t=100$~pA and $V_{\rm mod}=0.05$~mV$_{\rm rms}$.
(d)
High-resolution tunneling spectra at low $E$ taken at the same positions as for (c).
$V_s=+10$~mV, $I_t=100$~pA and $V_{\rm mod}=0.025$~mV$_{\rm rms}$.
Open symbols and solid lines denote experimental data and fitted results, respectively.
Spectra shown in (c) and (d) are shifted vertically for clarity.
(e)
The exponent $\alpha$ ($g(\bm{r}, E)\propto|E|^{\alpha}$) determined from the fit to the experimental data in the range of $|E| \leq 0.5$~meV plotted as a function of the distance from the TB.
}
\end{figure*}

We examined the LDOS evolution across the TB by taking a series of $g(\bm{r}, E)$ along the line indicated in Fig.~2(a).
Here and in the following, we are interested only in the evolution of $g(\bm{r}, E)$ along the $x$ axis running perpendicular to the TB leaving the $y$ coordinate constant, hence $g(x, E)$.
Figure~2(b) shows an intensity plot of $g(\bm{r}, E)$.
Individual spectra taken at representative points are depicted in Fig.~2(c).
At the position far away from the TB (I), $g(\bm{r}, E)$ exhibits a superconducting gap with clear quasiparticle peaks at $\approx \pm 2.5$~meV.
In addition to this main feature, there is a shoulder outside of the main peaks ($\approx \pm 3.5$~meV), which may represent multiple superconducting gaps~\cite{Kasahara2014PNAS}.
In contrast to the case of fully-gapped superconductors in which $g(\bm{r}, E)=0$ in an extended $E$ region near $E=0$, $g(\bm{r}, E)$ in FeSe approaches zero only for $E\rightarrow 0$ and apparently V-shaped, indicating the presence of line nodes~\cite{Song2011Science}.
Even right at the TB (III), the residual LDOS at $E=0$ is negligibly small, indicating that the TB hardly gives rise to a pair breaking effect.
In the vicinity of the TB, the quasiparticle peak and the shoulder associated with the superconducting gap diminish, and instead, sharp particle-hole symmetric peaks appear at $E \approx \pm 1.5$~meV.
In the crossover region (II), the 1.5~meV peak coexists with the 2.5~meV peak, meaning that the former is not associated with the suppressed superconducting gap.
The 1.5~meV peak diminishes within a distance of about 5~nm from the TB, which is close to the ``averaged'' in-plane coherence length $\xi_{ab}\approx 5$~nm obtained from the upper critical field $H_{c2}(\parallel c) \approx$15~T~\cite{Kasahara2014PNAS,Terashima2014PRB}.
These results suggest that the 1.5~meV peak represents the bound state induced by the TB.

Another interesting observation is that low-energy quasiparticle excitations are modified over a very long distance from the TB.
High-resolution $g(\bm{r}, E)$ spectra at the positions of (I), (II) and (III) are plotted in Fig.~2(d).
While the overall V-shaped behavior is maintained, the exact shape near the bottom of the gap depends on the position.
In order to examine this behavior, we fit an empirical power-law $g(\bm{r}, E) \propto |E|^\alpha$ to the low-energy ($|E|<0.5$~meV) spectra and plot the exponent $\alpha$ as a function of the distance from the TB at $x=0$ [Fig.~2(e)].
Except close to the TB ($|x|\lesssim\xi_{ab}$) where the 1.5~meV peaks dominate, $\alpha$ increases gradually with decreasing $x$ by about $\approx 40$\%.
This implies the suppression of the low-energy quasiparticle excitations, most probably due to the opening of a small gap induced by the TB.
The salient feature is that $\alpha$ continues to change even at $|x| > 10\xi_{ab}$ ($\approx 50$~nm), indicating an unexpectedly long-distance influence of the TB.

\begin{figure*}
\includegraphics[width=120mm]{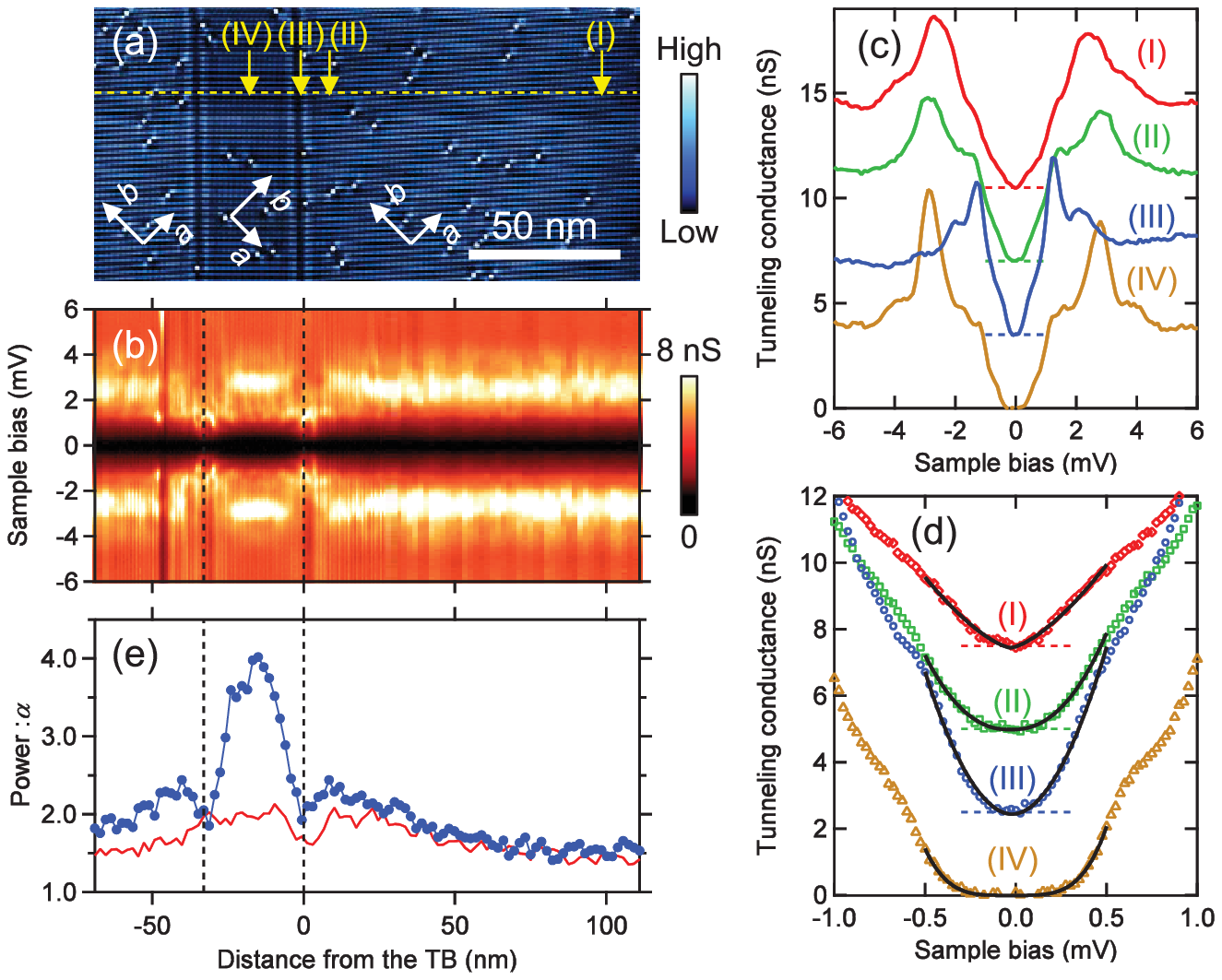}
\caption{
(a)
Constant-current STM image with double TBs taken at $V_s=+95$~mV and $I_t=10$~pA.
(b)
Intensity plot of $g(\bm{r}, E)$ along the yellow broken line in (a).
Positions of the TBs are indicated by broken lines.
A low-conductance position at $\approx$-47~nm is due to a point defect nearby.
$V_s=+20$~mV, $I_t=100$~pA and $V_{\rm mod}=0.05$~mV$_{\rm rms}$.
(c)
Tunneling spectra at the representative points I to IV indicated in (a).
$V_s=+20$~mV, $I_t=100$~pA and $V_{\rm mod}=0.05$~mV$_{\rm rms}$.
(d)
High-resolution tunneling spectra at low $E$ taken at the same positions as for (c).
$V_s=+10$~mV, $I_t=100$~pA and $V_{\rm mod}=0.025$~mV$_{\rm rms}$.
Symbols and solid lines denote experimental data and fitted results, respectively.
Spectra shown in (c) and (d) are shifted vertically for clarity.
(e)
The exponent, ($g(\bm{r}, E)\propto|E|^{\alpha}$) determined by the fitting in the range of $|E| \leq 0.5$~meV plotted as a function of the distance from one of the TB.
The data of the single TB is shown by a red line for reference.
}
\end{figure*}

The long-distance TB effect on the LDOS can be seen in a more dramatic way in two junctions in series formed by two TBs.
As shown in Fig.~3(a) we find an area where two TBs are running parallel to each other.
The distance between the TBs is 34~nm, which is about 7 times larger than $\xi_{ab}$.
Figure~3(b) shows the spatial evolution of $g(\bm{r}, E)$ across the double TBs.
Individual spectra at representative points are plotted in Fig.~3(c).
The overall spectral features, the 2.5~meV peak, the 3.5~meV shoulder and the 1.5~meV peak observed near a single TB are all reproduced (positions I, II, and III).
However, the low-energy spectrum taken inside the central domain (position IV) shows a striking anomaly which is absent in the case of a single TB.
Figure~3(d) depicts $g(\bm{r}, E)$ spectra at low energies.
It is clear that, in between the double TBs, there is a finite energy range where $g(\bm{r}, E)$ is almost completely zero.
The noticeable difference of the gap structure between inside and outside the central domain is clearly seen in Fig.~3(e), which shows the exponent $\alpha$ plotted as a function of the distance from one of the TBs; $\alpha$ is strongly enhanced in the central domain and peaks at the middle of the domain.
The large power $\alpha\approx4$, which is $\approx3$ times larger than the values at large $x$, essentially indistinguishable from an exponential energy dependence.
This apparent large power again corroborates the finite gap opening in the excitation spectrum of quasiparticle.

\subsection{Possible time-reversal-symmetry-broken state near the twin boundary}

\begin{figure*}
\includegraphics[width=110mm]{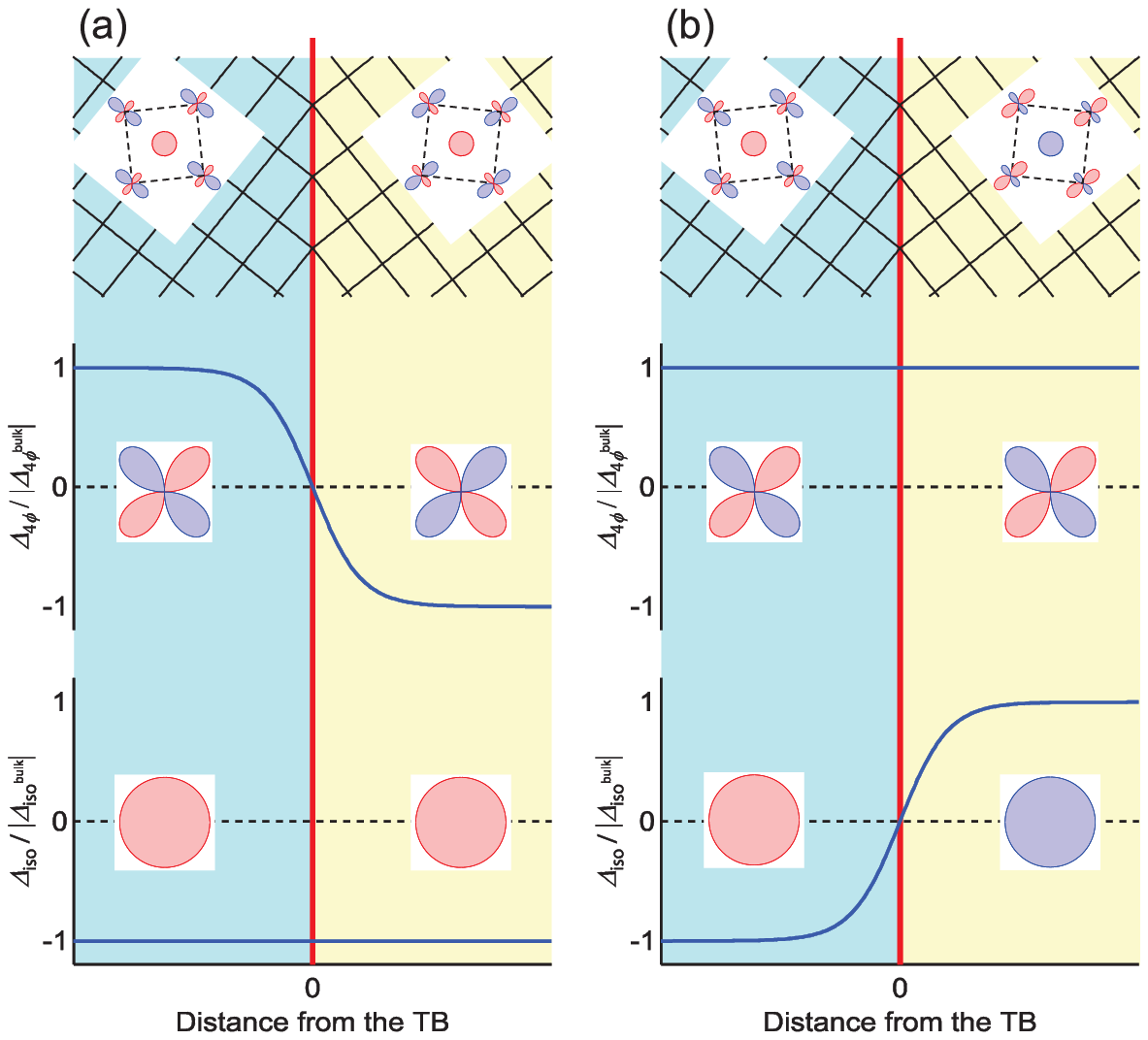}
\caption{
Schematic illustration of the phases of the superconducting gaps across the TB shown by the red line.
Top panel represents the iron lattice near the TB together with the momentum-space phase structure of the superconducting gaps opening at multiple Fermi cylinders,  a hole cylinder at the center and electron cylinders at the corner of the Brillouin zone (black broken square).
Different colors (red and blue) denote different signs of the phase.
We assume that the gap node exists on the electron cylinder and the sign reversal is between the main lobe of the gap on the electron cylinder and the gap on the hole cylinders but the argument given in the text applies not only for this particular case but also for other cases.
There are two possibilities: either the phase structure is fixed to the lattice (a) or is flipped across the TB (b).
In the former case, the nodal component $\Delta_{4\phi}$ should change its sign across the TB, whereas the sign of the isotropic component $\Delta_{\rm iso}$ (either due to the fully gapped Fermi cylinder or associated with the $C_2$ symmetry of the nodal gap) should  be reversed in the latter case.
}
\end{figure*}

The above observations, the TB-induced bound states at finite energies and the suppression of the low-energy quasiparticle excitations over a length scale much longer than $\xi_{ab}$, suggest a novel role of the TB in an unconventional superconductor.
Before discussing the origin of these anomalies, we briefly review what can be expected at a TB of FeSe.
Recent high-resolution laser-ARPES measurements of FeSe indicate that the hole cylinder is fully gapped~\cite{Okazaki_PC}, implying that the line nodes are present on the electron cylinder, as has been also inferred from vortex imaging~\cite{Song2011Science}.
Given this information, we consider two possible phase structures for symmetry of the superconducting gap across a TB as illustrated in Fig.~4, where either the global phase of the superconducting gap is fixed to the crystallographic axis [Fig.~4(a)] or is flipped across the TB [Fig.~4(b)].
It should be noted that the sign of either the nodal gap or the nodeless gap is reversed between the two domains in Fig.~4(a) or Fig.~4(b), respectively.
This means that the amplitude of at least one of the gaps vanishes at the TB, giving rise to the zero-energy quasiparticle state that should appear as a zero-energy peak in $g(\bm{r}, E)$.
This argument applies not only for the particular phase structure shown in Fig.~4 but also for a general case in which nodal and nodeless gaps reside on multiple Fermi surfaces.

The observed bound-state peak at 1.5~meV apparently contradicts this conjecture and suggests instead that the TB induces an additional gap component which shifts the position of a zero-energy peak to a finite energy.
We point out that, as long as the induced gap is real, a sum of the bulk gap and the TB-induced gap reverses its sign at a finite distance from the TB and still gives rise to a zero-energy peak.
However, as shown in Fig.~2(b), we did not observe a zero-energy peak in $g(\bm{r}, E)$ over more than 100~nm from the TB.
Thus, we speculate that the induced gap has an imaginary component, which means that time reversal symmetry is locally broken near the TB.
In such a case, bound state peaks are located at finite energies $E=\pm\Delta\cos(\delta\varphi/2)$ because the phase shift $\delta\varphi$ on the TB is reduced from $\pi$~\cite{Rainer1998JPCS,Furusaki1999SuperMicro}.
Here, $\Delta$ is the amplitude of the superconducting gap.
The possibility of the TB-induced time-reversal-symmetry-broken state has been argued theoretically in $d$-wave YBa$_2$Cu$_3$O$_{7-\delta}$ with a small $s$-wave component~\cite{Sigrist1996PRB}, but the experimental observation is still lacking.

\begin{figure*}
\includegraphics[width=160mm]{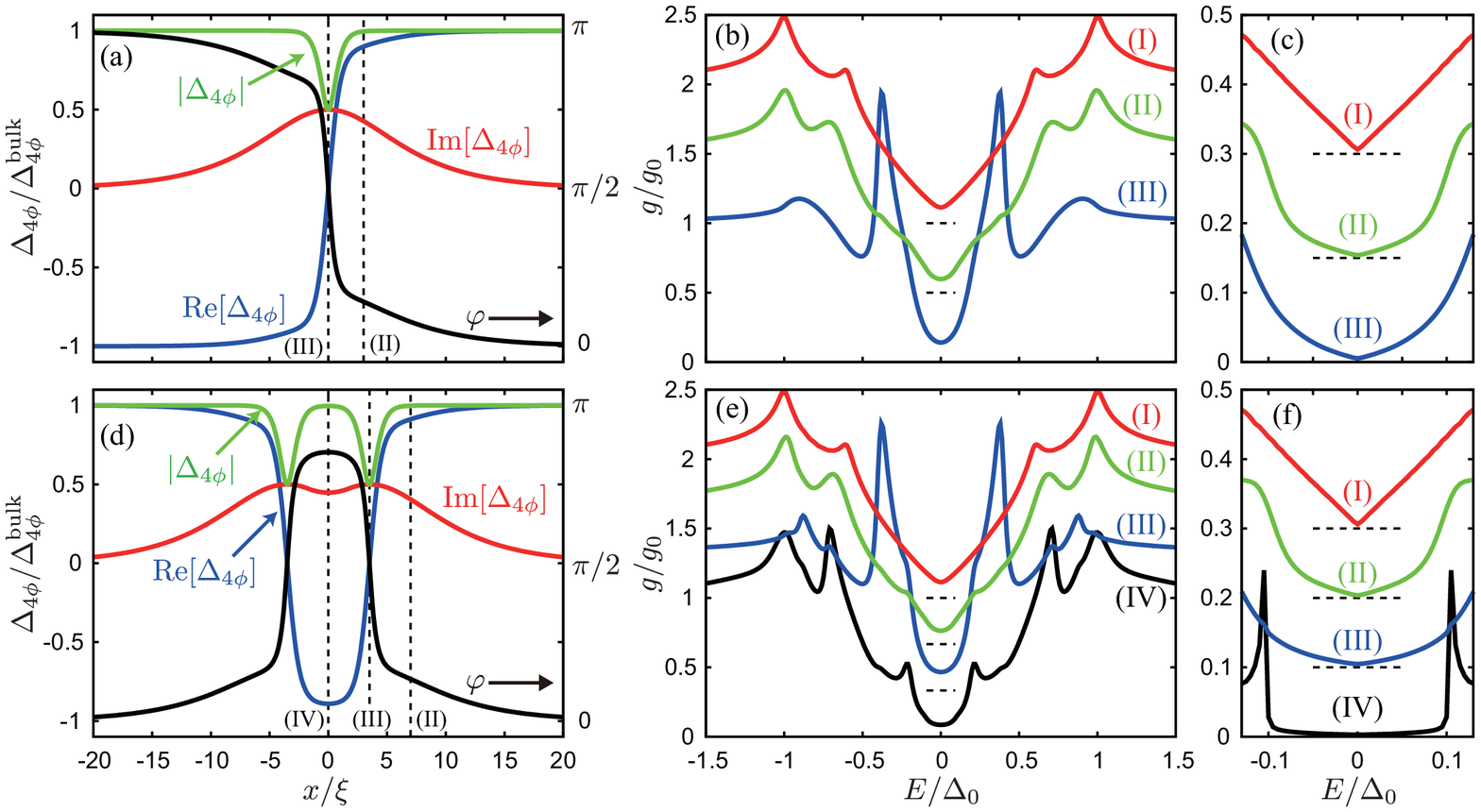}
\caption{
(a) A model order parameter $\Delta_{4\phi}(x)$ with a TB located at $x=0$.
LDOS's in the bulk (I), at $x=3\xi$ (II), and on a TB (III),
which are calculated with an energy smearing of $\eta=0.03\Delta_0$ (b)
and $\eta=0.001\Delta_0$ (c).
The lines (I) and (II) have offsets $g_0$ and $0.5g_0$ in (b)
and $0.3g_0$ and $0.15g_0$ in (c), respectively,
where $g_0$ is the density of states in the normal state at the Fermi energy.
(d) A model order parameter $\Delta_{4\phi}(x)$ with double TBs
located at $x=\pm 3.5\xi$.
The LDOS in the bulk (I), at $x=7\xi$ (II), on a TB (III),
and at the middle point between double TBs (IV),
which are calculated with $\eta=0.03\Delta_0$ (e) and $\eta=0.001\Delta_0$ (f).
The lines (I), (II), and (III) have offsets $g_0$, $(2/3)g_0$, and $(1/3)g_0$
in (e) and $0.3g_0$, $0.2g_0$, and $0.1g_0$, in (f) respectively.
}
\end{figure*}

In order to substantiate the relevance of this scenario, we have calculated the spatial evolution of the LDOS for a model order parameter with broken time-reversal symmetry near TBs.
The $C_2$-symmetric order parameter is represented by a sum of the isotropic component $\Delta_{\rm iso}$ and the four-fold nodal component $\Delta_{4\phi}\sin(2\phi)$,

\begin{equation}
\Delta(x)=\Delta_\mathrm{iso}+\Delta_{4\phi}(x)\sin(2\phi),
\end{equation}

\noindent
where $\phi$ is the azimuthal angle in the momentum space; see Fig.~4.
We assume that the global phase of the order parameter is fixed to the crystallographic axis, that is, the nodal component changes its sign across a TB as shown in Fig.~4(a).
The spatial variation of $\Delta_{4\phi}$ around a TB at $x=x_0$ is modeled by the form

\begin{equation}
\Delta_{4\phi}(x)
=\Delta_{4\phi}^{\rm bulk}
\{\tanh[(x-x_0)/\xi]\cos\theta(x)+i\sin\theta(x)\},
\end{equation}

\noindent
where the $x$ axis is taken to be perpendicular to the TB, and $\Delta_{4\phi}^{\rm bulk}$ is the amplitude of $\Delta_{4\phi}$ in the bulk.
The phase $\varphi$ of $\Delta_{4\phi}(x)$ equals $\theta(x)$ for $x-x_0\gg\xi$
and $\pi -\theta(x)$ for $-(x-x_0)\gg\xi$.
The phase variable $\theta(x)$ is assumed to take a nonvanishing value near the TB and exponentially decay with another length scale $\tilde\xi$.
It is important to note that the characteristic length $\tilde{\xi}$ for the local time-reversal symmetry breaking can be much longer than the coherence length $\xi$ \cite{Sigrist1996PRB}.
(The derivation of the length $\tilde{\xi}$ is given in Appendix~B.)
To account for low-energy excitations near the nodes observed in the LDOS, we focus on the electron cylinder with nodal gaps by setting the parameters $\Delta_{\rm iso}=0.2\Delta_0$ and $\Delta_{4\phi}^{\rm bulk}=0.8\Delta_0$.

As a model order parameter with a TB at $x=x_0=0$, we take $\theta(x)=(\pi/6){\rm sech}(x/\tilde{\xi})$ with $\tilde{\xi}=5\xi$, which gives $\Delta_{4\phi}(x=0)=(i/2)\Delta_{4\phi}^{\rm bulk}$.
The order parameter $\Delta_{4\phi}(x)$ is shown in Fig.~5(a), where $|\Delta_{4\phi}|$ changes with the length scale $\xi$ while ${\rm Im}(\Delta_{4\phi})$ decays with the longer length scale $\tilde{\xi}$.
The phase $\varphi$ abruptly changes near the TB and gradually approaches $0$ or $\pi$.
Using this order parameter, we calculate the spatial dependence of the LDOS within the quasi-classical approximation~\cite{Schopoh1995PRB}.
Figure 5(b) shows the global peak structure of the LDOS at representative points calculated with energy smearing of $\eta=0.03\Delta_0$.
Far from the TB, namely in the bulk (I), the LDOS has peaks at $|E|=\Delta_{4\phi}^{\rm bulk}\pm\Delta_{\rm iso}$.
On the TB (III), the peaks observed in the bulk are suppressed, and alternative peaks appear at $E\approx \pm0.4\Delta_0$, which correspond to the bound states whose energies are shifted from $E=0$ due to the local time-reversal symmetry breaking in $\Delta_{4\phi}$.
The bound-state peaks disappear at $x=3\xi$ (II) since their wave functions decay into the bulk with the length scale $\xi$.
These features of the calculated LDOS arising from the bound states are consistent with the LDOS peaks observed at $E\approx\pm 1.5$~meV by STM.
We show in Fig.~5(c) the LDOS at lower energy scale which has been calculated with a much smaller smearing factor $\eta=0.001\Delta_0$.
The clear V-shaped LDOS in the bulk (I) changes to the U-shaped LDOS upon approaching the TB, in agreement with the increase of the exponent $\alpha$ evaluated from the experimental data [Fig.~2(e)].
The low-energy LDOS is finite at $x=0$ (III) and $x=3\xi$ (II) because low-energy quasiparticles with momenta along the nodal directions of the bulk gap can linger over long distance and reach the TB, even though the local gap,

\begin{widetext}
\begin{equation}
|\Delta(x)|=\sqrt{[\Delta_{\rm iso}+{\rm Re}(\Delta_{4\phi})\sin(2\phi)]^2
+[{\rm Im}(\Delta_{4\phi})\sin(2\phi)]^2},
\end{equation}

\noindent
does not vanish near the TB where $\mathrm{Im}(\Delta_{4\phi})\ne0$.

We also calculate the LDOS for double TBs located at $x=\pm x_0=\pm 3.5\xi$, taking the model order parameter of the form

\begin{equation}
\Delta_{4\phi}(x)=\Delta_{4\phi}^{\rm bulk}\{\tanh[(x-x_0)/\xi]\tanh[(x+x_0)/\xi]\cos\theta(x)
+i\sin\theta(x)\},
\end{equation}
\end{widetext}

\noindent
shown in Fig.~5(d).
We assume that the distance $2x_0$ between the TBs is in the range $\xi\ll x_0\lesssim\tilde{\xi}$.
The phase $\theta(x)$ is an even function of $x$ and takes a maximum value at $x=0$.
The global peak structure of the LDOS and its low-energy blowup are shown for representative points along the $x$ direction in Fig.~5(e) and 5(f), respectively.
The large peaks at $|E|\approx 0.4\Delta_0$ on a TB (III) and the small peaks at $|E|\approx 0.2\Delta_0$ at the middle point $x=0$ between the TBs (IV) in Fig.~5(e) originate from the same dispersive mode of bound states at a TB.
The calculated LDOS spectrum at $x=0$ between the double TBs (IV) in Fig.~5(f) exhibits a clear energy gap extending over the region $|E|\lesssim 0.1\Delta_0$, reflecting the existence of a larger local gap at $x=0$, where the bulk low-energy quasiparticles cannot reach.
We conclude that the local gap enhanced by the local time-reversal symmetry breaking near TBs over the length scale $\tilde\xi$ can explain the strong suppression of the LDOS between the two TBs observed in our STM/STS experiments.

\section{Summary}

We have reported on the visualization of the atomic scale variation of the quasiparticle states of the nodal superconductor FeSe near TBs that enforce a sign inversion at least one of the superconducting gaps opening on multiple Fermi cylinders.
In contrast to the expectation that the sign inversion generates a zero-energy quasiparticle bound state near the TB, the TB-induced quasiparticle states are not at zero but at finite energies $E\approx\pm1.5$~meV.
Moreover, the low-energy excitation spectrum is affected by the TB over an extremely long distance, which is a few tens of times larger than $\xi_{ab}$.
An even more dramatic change in the low-energy spectrum has been detected in the region between double TBs separated by a distance $\approx 7\xi_{ab}$, where the quasiparticle weight near the Fermi energy is almost completely removed in the energy range $|E|\lesssim 0.2$~meV.
These observations are qualitatively reproduced by a phenomenological model which assumes that the TB induces locally a superconducting state that breaks time-reversal symmetry.

Our results suggest several important directions for future studies.
A microscopic mechanism that induces the time-reversal symmetry broken state is elusive and should be investigated theoretically.
It is also interesting to go beyond the quasi-classical approximation because FeSe is a unique superconductor whose Fermi energy is of the same order as the superconducting gap, placing this system in the BCS-BEC crossover regime~\cite{Kasahara2014PNAS}.
Experiments that directly probe the time-reversal symmetry breaking, such as muon-spin rotation and local magnetometry, are highly desirable and would give us further insights into the unconventional superconducting junctions.
We anticipate that the TBs in FeSe will stimulate further research on the role of the phase of the superconducting order parameter near the interface, which has been difficult to access experimentally.

\begin{acknowledgments}
This work has been supported by Japan -- Germany Research Cooperative Program, KAKENHI from JSPS and Project No. 56393598 from DAAD, and the ``Topological Quantum Phenomena" (No. 25103713) KAKENHI on Innovative Areas from MEXT of Japan.
\end{acknowledgments}

\appendix

\section{Absence of lattice distortion induced by the twin boundary}

\begin{figure*}
\includegraphics[width=130mm]{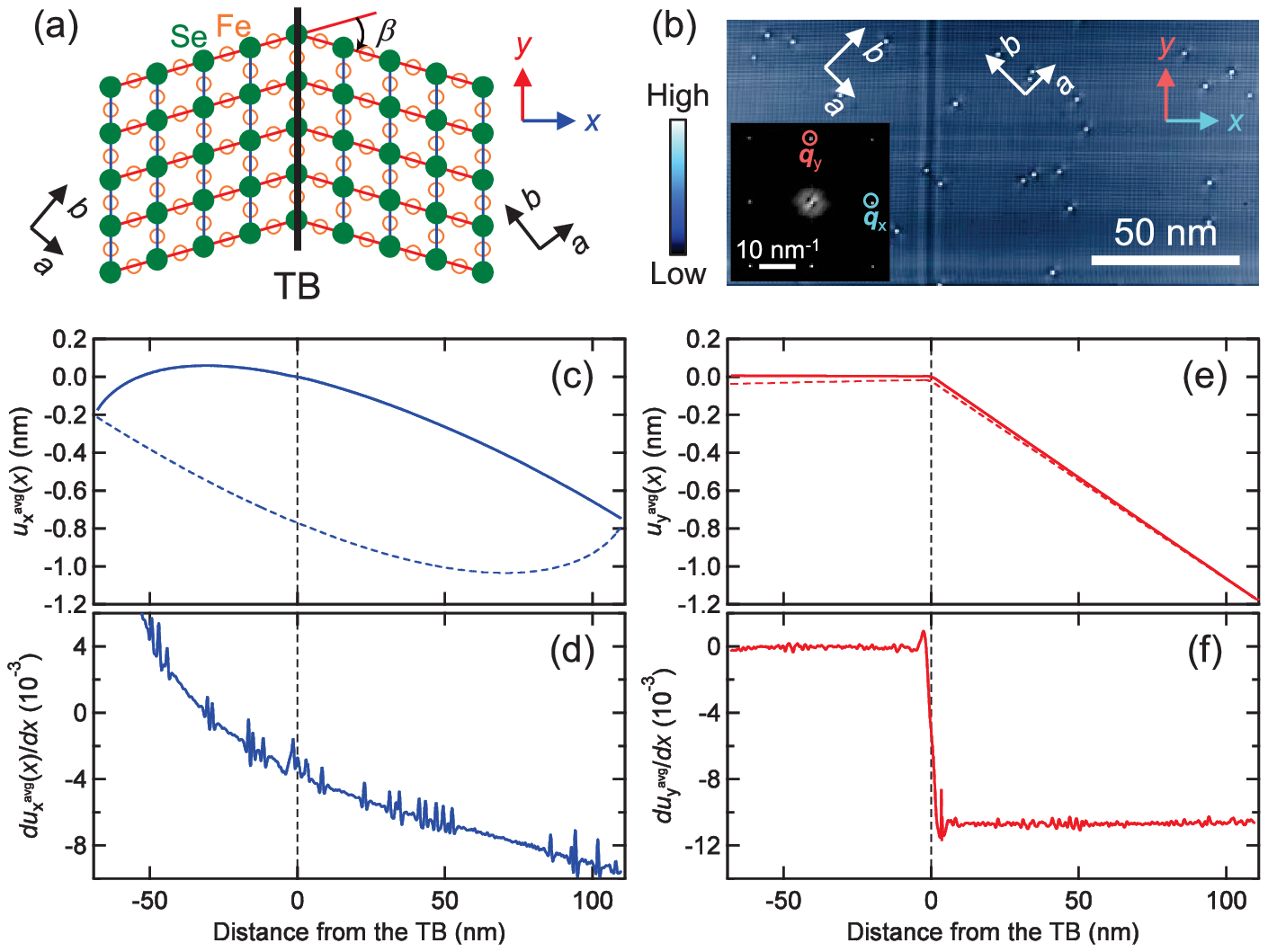}
\caption{
(a)
A schematic top view of the atomic arrangement near the TB of FeSe (not in scale).
Note that an atomic periodicity along the $x$ direction is hardly affected by the TB.
(b)
A constant-current STM image taken over a field of view of 180~nm$\times$90~nm on a grid of 4096$\times$2048 pixels.
The set-up conditions for imaging were $V_s=+95$~mV and $I_t=100$~pA.
Inset:
Fourier-transformed STM image taken in the left domain of the main panel.
A peak at $\bm{q}_x$ is sharp and well isolated from other features, guaranteeing that $\bm{q}_x\cdot\bm{u}(\bm{r})$ can be treated as a spatially varying phase of the $\bm{q}_x$-modulations.
(The same is true for $\bm{q}_y$.)
(c)
The $x$ component of $\bm{u}(\bm{r})$ averaged over the $y$ axis.
Thick solid line and thin dashed line denote the data taken by the forward (left to right) and backward (right to left) scans, respectively.
The symmetric hysteretic behavior between the forward and backward scans indicates that $u_x^\mathrm{avg}(x)$ is governed by the creep of the scanner.
No anomaly is observed at the TB.
(d)
The $x$ derivative of $u_x^\mathrm{avg}(x)$.
Spike-like features are associated with the point defects in the image.
(e)
The $y$ component of $\bm{u}(\bm{r})$ averaged over the $y$ axis.
Thick solid line and thin dashed line denote the data taken by the forward (left to right) and backward (right to left) scans, respectively.
Since the $y$ direction is the slow-scan direction, the effect of the creep is small.
A clear kink is observed at the TB.
(f)
The $x$ derivative of $u_y^\mathrm{avg}(x)$.
}
\end{figure*}

Although STM has a high spatial resolution, possible creep in the piezoelectric scanner and/or the thermal drift make it difficult to estimate the small distortions in the topographic image.
Here we utilize the so-called Lawler-Fujita algorithm~\cite{Lawler2010Nature} to deduce the lattice distortion and show that the TB-induced strain is negligibly small.

First we briefly explain the principle of the methodology.
The observed STM topography $T(\bm{r})$, which mainly represents the top-most Se lattice, can be expressed as

\begin{align}
&T(\bm{r})=\nonumber\\
&T_0\left[\cos\left\{\bm{q}_x\cdot\left(\bm{r}-\bm{u}(\bm{r})\right)\right\}+\cos\left\{\bm{q}_y\cdot\left(\bm{r}-\bm{u}(\bm{r})\right)\right\}\right]+\cdots.
\end{align}

\noindent
Here, $T_0$ is the amplitude of the Se-lattice modulation, $\bm{q}_x$ and $\bm{q}_y$ are wave vectors for the Se lattice, and $\cdots$ represent all other modulations.
The distortions from the perfect lattice is described by the displacement field $\bm{u}(\bm{r})$ that can be regarded as a spatially varying phase of the $\bm{q}_x$ and $\bm{q}_y$ modulations.
This approximation is justified as long as the length scale of distortions is much longer than the Se-Se distance $a_{\rm Se}$.
Standard phase-sensitive detection scheme can be used to evaluate $\bm{u}(\bm{r})$.
By multiplying $T(\bm{r})$ and the reference signal $\cos\left(\bm{q}_x\cdot\bm{r}\right)$, we get

\begin{align}
T(\bm{r})\cos\left(\bm{q}_x\cdot\bm{r}\right)=
\frac{T_0}{2}\Big[&\cos\left(\bm{q}_x\cdot\bm{u}(\bm{r})\right)\nonumber\\
&+\cos\left(2\bm{q}_x\cdot\bm{r}-\bm{q}_x\cdot\bm{u}(\bm{r})\right)\nonumber\\
&+\cos\left(\left(\bm{q}_x+\bm{q}_y\right)\cdot\bm{r}-\bm{q}_y\cdot\bm{u}(\bm{r})\right)\nonumber\\
&+\cos\left(\left(-\bm{q}_x+\bm{q}_y\right)\cdot\bm{r}-\bm{q}_y\cdot\bm{u}(\bm{r})\right)\Big]\nonumber\\
&+\cdots.
\end{align}

\noindent
All terms except the first exhibit periodic spatial modulations, which can be removed by low-pass Fourier filtering $\mathrm{LPF}\left\{\cdots\right\}$.

\begin{align}
\mathrm{LPF}\left\{T(\bm{r})\cos\left(\bm{q}_x\cdot\bm{r}\right)\right\}=\frac{T_0}{2}\cos\left(\bm{q}_x\cdot\bm{u}(\bm{r})\right).
\end{align}

\noindent
By using the quadrature reference $\sin\left(\bm{q}_x\cdot\bm{r}\right)$, we get

\begin{align}
\mathrm{LPF}\left\{T(\bm{r})\sin\left(\bm{q}_x\cdot\bm{r}\right)\right\}=\frac{T_0}{2}\sin\left(\bm{q}_x\cdot\bm{u}(\bm{r})\right).
\end{align}

\noindent
Therefore, we obtain $u_x(\bm{r})$, the $x$ component of $\bm{u}(\bm{r})$ as
\begin{align}
u_x(\bm{r})=\frac{a_{\rm Se}}{2\pi}\tan^{-1}\frac{\mathrm{LPF}\left\{T(\bm{r})\sin\left(\bm{q}_x\cdot\bm{r}\right)\right\}}{\mathrm{LPF}\left\{T(\bm{r})\cos\left(\bm{q}_x\cdot\bm{r}\right)\right\}}.
\end{align}

\noindent
The $y$ component $u_y(\bm{r})$ can also be deduced as

\begin{align}
u_y(\bm{r})=\frac{a_{\rm Se}}{2\pi}\tan^{-1}\frac{\mathrm{LPF}\left\{T(\bm{r})\sin\left(\bm{q}_y\cdot\bm{r}\right)\right\}}{\mathrm{LPF}\left\{T(\bm{r})\cos\left(\bm{q}_y\cdot\bm{r}\right)\right\}}.
\end{align}

A schematic model of atomic arrangement near the TB is shown in Fig.~6(a).
We expect that the orthorhombic distortion affects the atomic arrangement along the $y$ direction across the TB, while the periodicity along the $x$ direction remains intact.
In order to verify this model and to check if there is an additional lattice distortion, we calculated $u_x(\bm{r})$ and $u_y(\bm{r})$ of the high-resolution STM image containing a TB running along the $y$ direction [Fig~6(b)].
Reference wave vectors $\bm{q}_x$ and $\bm{q}_y$ were obtained by the Fourier analysis in the left domain.
For low-pass Fourier filtering, we picked up only long-wavelength components by using a Gaussian mask with half width at the half maximum of $0.21(2\pi/a_{\rm Se})$.
Since there is a translational symmetry along the TB, we average $u_x(\bm{r})$ and $u_y(\bm{r})$ along the $y$ direction, yielding $u_x^\mathrm{avg}(x)$ and $u_y^\mathrm{avg}(x)$, respectively.
This significantly enhances the signal-to-noise ratio.

Figures~6(c) and (d) show $u_x^\mathrm{avg}(x)$ and its $x$ derivative.
There is no noticeable anomaly in both $u_x^\mathrm{avg}(x)$ and $du_x^\mathrm{avg}(x)/dx$, except for the smooth background associated with the creep of the piezoelectric scanner.
By contrast, $u_y^\mathrm{avg}(x)$ exhibits a sharp kink at the TB [Fig.~6(e)].
These features are consistent with the model shown in Fig.~6(a).
It should be noted that $du_y^\mathrm{avg}(x)/dx$ shown in Fig.~6(f) is almost completely constant in both domains, indicating that the TB-induced strain to the lattice is negligibly small.

The observed value of $du_y^\mathrm{avg}(x)/dx \approx -1.1\times10^{-2}$ in the right domain means that the angle $\beta$ defined in Fig.~6(a) is +0.63$^\circ$.
This means that orthorhombic distortion $(b-a)/(b+a) \approx 2.8\times10^{-3}$, being consistent with the X-ray result~\cite{Khasanov2010NJP}.
Even if there were an additional lattice distortion associated with the TB, it should be much smaller than this tiny orthorhombic distortion which we have clearly detected.

\section{Asymptotic forms of the order parameter derived by the Ginzburg-Landau theory}

We derive asymptotic forms of the order parameter far from TBs using the Ginzburg-Landau (GL) theory.
We consider the GL free-energy functional for tetragonal symmetric systems~\cite{Sigrist1996PRB} as an expansion in the isotropic $s$-wave component $\Delta_{\rm iso}$ and the four-fold $d$-wave component $\Delta_{4\phi}$ of the order parameter:
\begin{widetext}
\begin{multline}
F_{\rm GL}[\Delta_{\rm iso},\Delta_{4\phi}]
=\int dV\left\{\sum_{\mu={\rm iso},4\phi}
\left[
\tilde{a}_{\mu}(T)|\Delta_{\mu}|^2
+b_{\mu}|\Delta_{\mu}|^4
+K_{\mu}|\bm{\nabla}\Delta_{\mu}|^2\right]
+\gamma_1|\Delta_{\rm iso}|^2|\Delta_{4\phi}|^2 \right.\\
\left.
+\frac{\gamma_2}{2}\left(\Delta_{\rm iso}^{*2}\Delta_{4\phi}^2
+\Delta_{\rm iso}^2\Delta_{4\phi}^{*2}\right)
+\frac{\widetilde{K}}{2}\left[
(\partial_a\Delta_{\rm iso})^*(\partial_a\Delta_{4\phi})
-(\partial_b\Delta_{\rm iso})^*(\partial_b\Delta_{4\phi})
+{\rm c.c.}\right]\right\},
\end{multline}
\end{widetext}
where we have neglected the vector potential as it does not play an important role in our discussion.
The coefficients $b_{\mu }$, $K_{\mu }$, and $\widetilde{K}$ are positive and $\tilde{a}_{\mu }(T)=a_{\mu }(T/T_{\rm c\mu }-1)$ with positive $a_{\mu}$.
The differential operator $\bm{\nabla}=(\partial_a,\partial_b)$ is defined according to the crystal axes $a$ and $b$.
As in Ref.~\onlinecite{Sigrist1996PRB}, we assume $\gamma_2>0$, so that the free energy is minimized at $\varphi=\pm\pi/2$, and the time-reversal-symmetry-broken $s\pm id$ state is stabilized when both $\Delta_{\rm iso}$ and $\Delta_{4\phi}$ are finite.

The effect of orthorhombic distortion is taken into account by adding the following term to the free-energy functional \cite{Sigrist1996PRB}:
\begin{align}
F_{\epsilon }=
c\epsilon\int dV(\Delta_{\rm iso}^*\Delta_{4\phi}
+\Delta_{\rm iso}\Delta_{4\phi}^*),
\end{align}
where $c$ is a positive parameter and $\epsilon=\epsilon_{aa}-\epsilon_{bb}$ is the parameter of the orthorhombic lattice distortion.
The total free energy for a uniform state in the bulk is then given by
\begin{widetext}
\begin{align}
\frac{F_{\rm GL}+F_{\epsilon }}{V}=&
\sum_{\mu={\rm iso},4\phi}
\left(\tilde{a}_{\mu}|\Delta_{\mu}|^2+b_{\mu}|\Delta_{\mu}|^4\right)\nonumber\\
&+\gamma_1|\Delta_{\rm iso}|^2|\Delta_{4\phi}|^2
+\gamma_2|\Delta_{\rm iso}|^2|\Delta_{4\phi}|^2\cos(2\varphi)
+2c\epsilon|\Delta_{\rm iso}||\Delta_{4\phi}|\cos\varphi,
\end{align}
where $\Delta_{\mu}=|\Delta_{\mu}|e^{i\varphi_{\mu }}$
and the relative phase $\varphi=\varphi_{4\phi}-\varphi_{\rm iso}$.
If $c|\epsilon|\ge 2\gamma_2|\Delta_{\rm iso}||\Delta_{4\phi}|$,
then the free energy is minimized at $\varphi=0$ for $\epsilon<0$
and at $\varphi=\pi$ for $\epsilon>0$.
In the following discussion we assume that this inequality is satisfied and the time reversal symmetric $s\pm d$ state is realized in the bulk.

Next, we consider a TB located at $x=x_0$ along the $y$ axis, where the $x$ and $y$ axes are rotated by 45$^\circ$ from the crystalline axes, $x=(a-b)/\sqrt{2}$ and $y=(a+b)/\sqrt{2}$.
The orthorhombic lattice distortion parameter $\epsilon$ changes its sign across the TB.
We assume $\epsilon(x)\to\mp|\epsilon|$ for $x\to\pm\infty$, so that the $s\pm d$ state is realized in $x\to\pm\infty$.
Near the TB where $\epsilon(x)$ is small, the $s\pm id$ state is favored.
Then, the area density of the total free energy is given by
\begin{align}
f_{\rm GL}+f_{\epsilon}=
\int dx\Bigg[&
\sum_{\mu={\rm iso},4\phi}\left(\tilde{a}_{\mu}|\Delta_{\mu}|^2
+b_{\mu}|\Delta_{\mu}|^4+K_{\mu}|\partial_x\Delta_{\mu}|^2\right)
+\gamma_1|\Delta_{\rm iso}|^2|\Delta_{4\phi}|^2
\nonumber\\
&
+\frac{\gamma_2}{2}\left(\Delta_{\rm iso}^{*2}\Delta_{4\phi}^2
+\Delta_{\rm iso}^2\Delta_{4\phi}^{*2}\right)
+c\epsilon(x)(\Delta_{\rm iso}^*\Delta_{4\phi}
+\Delta_{\rm iso}\Delta_{4\phi}^*)\Bigg].
\end{align}
Let us first assume that $\Delta_{\rm iso}$ is a real and uniform order parameter while $\Delta_{4\phi}$ changes its sign across the TB, as shown in Fig.~4(a).
If we restrict $\Delta_{4\phi}$ to be real, then $\Delta_{4\phi}$ varies over the coherence length~\cite{Sigrist1996PRB}
\begin{align}
\xi=\sqrt{
\frac{K_{4\phi}}
{\tilde{a}_{4\phi}+6b_{4\phi}|\Delta_{4\phi}^{\rm bulk}|^2
+(\gamma_1+\gamma_2)|\Delta_{\rm iso}|^2}
},
\end{align}
where $|\Delta_{4\phi}^{\rm bulk}|$ is the amplitude of $\Delta_{4\phi}$ in the bulk.
However, we expect that time-reversal symmetry should be locally broken at the TB.
Thus, we allow $\Delta_{4\phi}$ to be complex, $\Delta_{4\phi}(x)=|\Delta_{4\phi}(x)|e^{i\varphi(x)}$.
With this order parameter, the total free energy is given by
\begin{align}
f_{\rm GL}+f_{\epsilon}=
\int \!dx\Bigg[&
\tilde{a}_{\rm iso}|\Delta_{\rm iso}|^2+b_{\rm iso}|\Delta_{\rm iso}|^4
+\tilde{a}_{4\phi}|\Delta_{4\phi}|^2+b_{4\phi}|\Delta_{4\phi}|^4
+\gamma_1|\Delta_{\rm iso}|^2|\Delta_{4\phi}|^2 \nonumber\\
&
+\gamma_2|\Delta_{\rm iso}|^2|\Delta_{4\phi}|^2\cos(2\varphi)
+2c\epsilon(x-x_0)|\Delta_{\rm iso}||\Delta_{4\phi}|\cos\varphi \nonumber\\
&
+K_{4\phi}\left[(\partial_x|\Delta_{4\phi}|)^2
 +|\Delta_{4\phi}|^2(\partial_x\varphi)^2\right]\Bigg].
\end{align}
\end{widetext}
In the bulk region ($x-x_0\gg\xi$) where $\partial_x|\Delta_{4\phi}|=0$
and $\epsilon(x)=-|\epsilon|$, the GL differential equation
to minimize $f_{\rm GL}+f_{\epsilon}$ is
\begin{align}
K_{4\phi}\partial_x^2\varphi=
-\frac{|\Delta_{\rm iso}|}{|\Delta_{4\phi}^{\rm bulk}|}
\left[\gamma_2|\Delta_{\rm iso}||\Delta_{4\phi}^{\rm bulk}|\sin(2\varphi)
     -c|\epsilon|\sin\varphi\right].
\end{align}
Since $\varphi\ll 1$ far away from the TB, we can linearize the differential equation and find the relative phase to decay as $\varphi\propto \exp\left(-x/\tilde{\xi}\right)$ with the characteristic length
\begin{align}
\tilde{\xi}=
\sqrt{
\frac{K_{4\phi}|\Delta_{4\phi}^{\rm bulk}|}
{|\Delta_{\rm iso}|
 (c|\epsilon|-2\gamma_2|\Delta_{\rm iso}||\Delta_{4\phi}^{\rm bulk}|)}
}.
\end{align}
The characteristic length diverges when approaching the phase boundary, where $c|\epsilon|=2\gamma_2|\Delta_{\rm iso}||\Delta_{4\phi}|$, between the time reversal symmetric $s\pm d$ state and the time-reversal-symmetry-broken $s\pm id$ state.

Finally, we consider double TBs at $x=\pm|x_0|$, where $\xi\ll |x_0|\lesssim\tilde{\xi }$.
We assume $\epsilon>0$ between the double TBs and $\epsilon<0$ otherwise.
At the center $x=0$ between the double TBs, we can set $|\Delta_{4\phi}|=|\Delta_{4\phi}^{\rm bulk}|$ because $|x_0|\gg\xi$.
With this approximation, the GL differential equation to minimize $f_{\rm GL}+f_{\epsilon}$ for $|x|\ll|x_0|$ is
\begin{align}
K_{4\phi}\partial_x^2\varphi=
-\frac{|\Delta_{\rm iso}|}{|\Delta_{4\phi}^{\rm bulk}|}
\left[\gamma_2|\Delta_{\rm iso}||\Delta_{4\phi}^{\rm bulk}|\sin(2\varphi)
+c|\epsilon|\sin\varphi\right].
\end{align}
Integration of the differential equation yields
\begin{widetext}
\begin{align}
K_{4\phi}(\partial_x\varphi)^2=
\frac{|\Delta_{\rm iso}|}{|\Delta_{4\phi}^{\rm bulk}|}
[\gamma_2|\Delta_{\rm iso}||\Delta_{4\phi}^{\rm bulk}|\cos(2\varphi)
+2c|\epsilon|\cos\varphi
-\gamma_2|\Delta_{\rm iso}||\Delta_{4\phi}^{\rm bulk}|\cos(2\varphi_0)
-2c|\epsilon|\cos\varphi_0],
\label{eq:differential}
\end{align}
\end{widetext}
where the integration constant is determined from the conditions $\partial_x\varphi(x=0)=0$ and $\varphi(x=0)\equiv\varphi_0$.
Since we assume the distance between the TBs is in the range $|x_0|\lesssim\tilde{\xi}$, the relative phase does not reach $\pi$ at $x=0$, i.e., $\varphi_0<\pi$.
For $\varphi_0-\varphi\ll1$ near $x=0$, the differential equation \eqref{eq:differential} has the solution
\begin{align}
\varphi(x)=\varphi_0-\left(\frac{x}{\tilde{\xi}_0}\right)^2.
\label{eq:varphi_dTB}
\end{align}
For the model order parameter shown in Fig.~5(d), we have determined $\varphi_0$ and $\tilde{\xi}_0$ by the continuity condition at $x=\pm|x_0|/2$, that is, by imposing that $\varphi=\pi -(\pi/6){\rm sech}(x/\tilde{\xi})$ and $\varphi(x)$ in Eq.~\eqref{eq:varphi_dTB} are smoothly connected at $x=\pm|x_0|/2$.
We note that different choices of the connecting position yield little change in the value of $\varphi$.


\begin{thebibliography}{99}
\bibitem{Duzer_book} T. Van Duzer and C. W. Turner, \textit{Principles of Superconductive Devices and Circuits} (Prentice Hall, New Jersey, 1998), 2nd ed.
\bibitem{Tsui2000RMP} C. C. Tsuei and J. R. Kirtley, \textit{Pairing symmetry in cuprate superconductors}, Rev. Mod. Phys. \textbf{72}, 969 (2000).
\bibitem{Kim2012NatPhys} J. Kim, V. Chua, G. A. Fiete, H. Nam, A. H. MacDonald, and C. -K. Shih \textit{Visualization of geometric influences on proximity effects in heterogeneous superconductor thin films} Nature Phys. \textbf{8}, 464 (2012).
\bibitem{Serrier-Garcia2013PRL} L. Serrier-Garcia, J. C. Cuevas, T. Cren, C. Brun, V. Cherkez, F. Debontridder, D. Fokin, F. S. Bergeret, and D. Roditchev \textit{Scanning Tunneling Spectroscopy Study of the Proximity Effect in a Disordered Two-Dimensional Metal}, Phys. Rev. Lett. \textbf{110}, 157003 (2013).
\bibitem{Cherkez2014PRX} V. Cherkez, J. C. Cuevas, C. Brun, T. Cren, G. Menard, F. Debontridder, V. S. Stolyarov, and D. Roditchev, \textit{Proximity Effect between Two Superconductors Spatially Resolved by Scanning Tunneling Spectroscopy}, Phys. Rev. X \textbf{4}, 011033 (2014).
\bibitem{Song2011Science} C. -L. Song, Y. -L. Wang, P. Cheng, Y. -P. Jiang, W. Li, T. Zhang, Z. Li, K. He, L. Wang, J. -F. Jia, H. -H Hung, C. Wu, X. Ma, X. Chen, Q. -K. Xue, \textit{Direct Observation of Nodes and Twofold Symmetry in FeSe Superconductor}, Science \textbf{332}, 1410 (2011).
\bibitem{Kasahara2014PNAS} S. Kasahara, T. Watashige, T. Hanaguri, Y. Kohsaka, T. Yamashita, Y. Shimoyama, Y. Mizukami, R. Endo, H. Ikeda, K. Aoyama, T. Terashima, S. Uji, T. Wolf, H. v. L\"{o}hneysen, T. Shibauchi, and Y. Matsuda, \textit{Field-induced superconducting phase of FeSe in the BCS-BEC cross-over}, Proc. Natl. Acad. Sci. \textbf{111}, 16309 (2014).
\bibitem{Derro2000PhysicaC} D. J. Derro, S. H. Pan, E. W. Hudson, K.M. Lang, J. C. Davis, K. Mochizuki, J. T. Markert, and A. de Lozanne, \textit{A Detailed Scanning Tunneling Microscopy Study of the CuO chains of YBa$_2$Cu$_3$O$_{7-x}$}, Physica C \textbf{341-348}, 425 (2000).
\bibitem{Chuang2010Science} T. -M. Chuang, M. P. Allan, Jinho Lee, Yang Xie, Ni Ni, S. L. Bud'ko, G. S. Boebinger, P. C. Canfield, and J. C. Davis, \textit{Nematic Electronic Structure in the "Parent" State of the Iron-Based Superconductor Ca(Fe$_{1-x}$Co$_x$)$_2$As$_2$}, Science \textbf{327}, 181 (2010).
\bibitem{Rosenthal2014NatPhys} E. P. Rosenthal, E. F. Andrade, C. J. Arguello, R. M. Fernandes, L. Y. Xing, X. C.Wang, C. Q. Jin, A. J. Millis, and A. N. Pasupathy, \textit{Visualization of electron nematicity and unidirectional antiferroic fluctuations at high temperatures in NaFeAs}, Nature Phys. \textbf{10}, 225 (2014).
\bibitem{Hsu2008PNAS} F. -C. Hsu, J. -Y. Luo, K. -W. Yeh, T. -K. Chen, T. -W. Huang, P. M. Wu, Y. -C. Lee, Y. -L. Huang, Y. -Y. Chu, D. -C. Yan, and M. -K. Wu, \textit{Superconductivity in the PbO-type structure $\alpha$-FeSe}, Proc. Natl. Acad. Sci. \textbf{105}, 14262 (2008).
\bibitem{Nakayama2014} K. Nakayama, Y. Miyata, G. N. Phan, T. Sato, Y. Tanabe, T. Urata, K. Tanigaki, and T. Takahashi, \textit{Reconstruction of Band Structure Induced by Electronic Nematicity in an FeSe Superconductor}, Phys. Rev. Lett. \textbf{113}, 237001 (2014).
\bibitem{Shimojima2014PRB} T. Shimojima, Y. Suzuki, T. Sonobe, A. Nakamura, M. Sakano, J. Omachi, K. Yoshioka, M. Kuwata-Gonokami, K. Ono, H. Kumigashira, A. E. B\"{o}hmer, F. Hardy, T. Wolf, C. Meingast, H. v. L\"{o}hneysen, H. Ikeda, and K. Ishizaka, \textit{Lifting of xz/yz orbital degeneracy at the structural transition in detwinned FeSe}, Phys. Rev. B \textbf{90}, 121111(R) (2014).
\bibitem{Watson2015condmat_a} M. D. Watson, T. K. Kim, A. A. Haghighirad, N. R. Davies, A. McCollam, A. Narayanan, S. F. Blake, Y. L. Chen, S. Ghannadzadeh, A. J. Schofield, M. Hoesch, C. Meingast, T. Wolf, and A. I. Coldea, \textit{Emergence of the nematic electronic state in FeSe}, arXiv:1502.02917 (2015).
\bibitem{Zhang2015condmat} P. Zhang, T. Qian, P. Richard, X. P. Wang, H. Miao, B. Q. Lv, B. B. Fu, T. Wolf, C. Meingast, X. X. Wu, Z. Q. Wang, J. P. Hu, and H. Ding, \textit{Evidence for intertwining orders in the electronic nematic state of FeSe}, arXiv:1503.01391 (2015).
\bibitem{Boehmer2015PRL} A. E. B\"{o}hmer, T. Arai, F. Hardy, T. Hattori, T. Iye, T. Wolf, H. v. L\"{o}hneysen, K. Ishida, and C. Meingast, \textit{Origin of the tetragonal-to-orthorhombic phase transition in FeSe: A combined thermodynamic and NMR study of nematicity}, Phys. Rev. Lett. \textbf{114}, 027001 (2015).
\bibitem{Baek2015NatMat} S-H. Baek, D. V. Efremov, J. M. Ok, J. S. Kim, Jeroen van den Brink, and B. B\"{u}chner, \textit{Orbital-driven nematicity in FeSe}, Nature Mat. \textbf{14}, 210 (2015).
\bibitem{Subedi2008PRB} A. Subedi, L. Zhang, D. J. Singh, and M. H. Du, \textit{Density functional study of FeS, FeSe, and FeTe: Electronic structure, magnetism, phonons, and superconductivity}, Phys. Rev. B \textbf{78}, 134514 (2008).
\bibitem{Aichhorn2010PRB}M. Aichhorn, S. Biermann, T. Miyake, A. Georges, and M. Imada, \textit{Theoretical evidence for strong correlations and incoherent metallic state in FeSe}, Phys. Rev. B \textbf{82}, 064504 (2010).
\bibitem{Terashima2014PRB} T. Terashima, N. Kikugawa, A. Kiswandhi, E. -S. Choi, J. S. Brooks, S. Kasahara, T. Watashige, H. Ikeda, T. Shibauchi, Y. Matsuda, T. Wolf, A. E. B\"{o}hmer, F. Hardy, C. Meingast, H. v. L\"{o}hneysen, M. Suzuki, R. Arita, and S. Uji, \textit{Anomalous Fermi surface in FeSe seen by Shubnikov-de Haas oscillation measurements}, Phys. Rev. B \textbf{90}, 144517 (2014).
\bibitem{Watson2015condmat_b} M. D. Watson, T. Yamashita, S. Kasahara, W. Knafo, M. Nardone, J. Beard, F. Hardy, A. McCollam, A. Narayanan, S. F. Blake, T. Wolf, A. A. Haghighirad, C. Meingast, A. J. Schofield, H. v. L\"{o}hneysen, Y. Matsuda, A. I. Coldea, and T. Shibauchi, \textit{Dichotomy between the hole and electrons behavior in the multiband FeSe probed by ultra high magnetic fields }, arXiv:1502.02922 (2015).
\bibitem{Maletz2014PRB} J. Maletz, V. B. Zabolotnyy, D. V. Evtushinsky, S. Thirupathaiah, A. U. B. Wolter, L. Harnagea, A. N. Yaresko, A. N. Vasiliev, D. A. Chareev, A. E. B\"{o}hmer, F. Hardy, T. Wolf, C. Meingast, E. D. L. Rienks, B. B\"{u}chner, and S. V. Borisenko, \textit{Unusual band renormalization in the simplest iron-based superconductor FeSe$_{1-x}$}, Phys. Rev. B \textbf{89}, 220506(R) (2014).
\bibitem{Song2012PRL} C. -L. Song, Y.-L. Wang, Y. -P. Jiang, L. Wang, K. He, X. Chen, J. E. Hoffman, X. -C. Ma, and Q.-K. Xue, \textit{Suppression of Superconductivity by Twin Boundaries in FeSe}, Phys. Rev. Lett. \textbf{109}, 137004 (2012).
\bibitem{Hanaguri2006JPhys} T. Hanaguri, \textit{Development of high-field STM and its application to the study on magnetically-tuned criticality in Sr$_3$Ru$_2$O$_7$}, J. Phys.: Conference Series \textbf{51}, 514 (2006).
\bibitem{Monma2011JAC} K. Momma and F. Izumi, \textit{VESTA 3 for three-dimensional visualization of crystal, volumetric and morphology data}, J. Appl. Cryst. \textbf{44}, 1272 (2011).
\bibitem{Boehmer2013PRB} A. E. B\"{o}hmer, F. Hardy, F. Eilers, D. Ernst, P. Adelmann, P. Schweiss, T. Wolf, and C. Meingast, \textit{Lack of coupling between superconductivity and orthorhombic distortion in stoichiometric single-crystalline FeSe}, Phys. Rev. B \textbf{87}, 180505(R) (2013).
\bibitem{Okazaki_PC} K. Okazaki, private communication.
\bibitem{Rainer1998JPCS} D. Rainer, H. Burkhardt, M. Fogelstr\"{o}m, and J. A. Sauls, \textit{Andreev bound states, surfaces and subdominant pairing in high $T_c$ superconductors}, J. Phys. Chem. Solids \textbf{59}, 2040 (1998).
\bibitem{Furusaki1999SuperMicro} A. Furusaki, \textit{Josephson current carried by Andreev levels in superconducting quantum point contacts}, Superlattices and Microstructures, \textbf{25}, 809 (1999).
\bibitem{Sigrist1996PRB} M. Sigrist, K. Kuboki, P. A. Lee, A. J. Millis, and T. M. Rice, \textit{Influence of twin boundaries on Josephson junctions between high-temperature and conventional superconductors}, Phys. Rev. B \textbf{53}, 2835 (1996).
\bibitem{Schopoh1995PRB} N. Schopohl and K. Maki, \textit{Quasiparticle spectrum around a vortex line in a d-wave superconductor}, Phys. Rev. B \textbf{52}, 490 (1995).
\bibitem{Lawler2010Nature} M. J. Lawler, K. Fujita, Jhinhwan Lee, A. R. Schmidt, Y. Kohsaka, Chung Koo Kim, H. Eisaki, S. Uchida, J. C. Davis, J. P. Sethna and Eun-Ah Kim, \textit{Intra-unit-cell electronic nematicity of the high-T$_c$ copper-oxide pseudogap states}, Nature \textbf{466}, 347 (2010).
\bibitem{Khasanov2010NJP}R. Khasanov, M. Bendele, K. Conder, H. Keller, E. Pomjakushina and V. Pomjakushin, \textit{Iron isotope effect on the superconducting transition temperature and the crystal structure of FeSe$_{1-x}$} New J. Phys., \textbf{12}, 073024 (2010).
\end{thebibliography}
\end{document}